\documentclass[twocolumn,showpacs,amsmath,amssymb,prl,superscriptaddress]{revtex4}

\usepackage{graphicx}
\usepackage{epstopdf}
\usepackage{bm}

\begin{document}

\title{BCS superconductivity of Dirac electrons in graphene layers}
\author{N.\ B. Kopnin}
\affiliation{Low Temperature Laboratory, Helsinki University of
Technology, P.O. Box 2200, FIN-02015 HUT, Finland} \affiliation{
L. D. Landau Institute for Theoretical Physics, 117940 Moscow,
Russia}
\author{E.\ B. Sonin}
\affiliation{The Racah Institute of Physics, Hebrew University of
Jerusalem, Israel}

\date{\today}

\begin{abstract}
Possible superconductivity of electrons with the Dirac spectrum is
analyzed using  the BCS model. We calculate the critical
temperature, the superconducting energy gap, and supercurrent as
functions of the doping level and of the pairing interaction
strength. Zero doping is characterized by existence of the quantum
critical point such that the critical temperature vanishes below
some finite value of the interaction strength. However, the
critical temperature remains finite for any nonzero electron or
hole doping level when the Fermi energy is shifted away from the
Dirac point of the normal-state electron spectrum. We analyze the
behavior of the characteristic length scales, i.e., the London
penetration depth and the coherence length, which determine the
critical magnetic fields.
\end{abstract}

\maketitle

\begin{figure}[t]
\centerline{\includegraphics[width=0.9\linewidth]{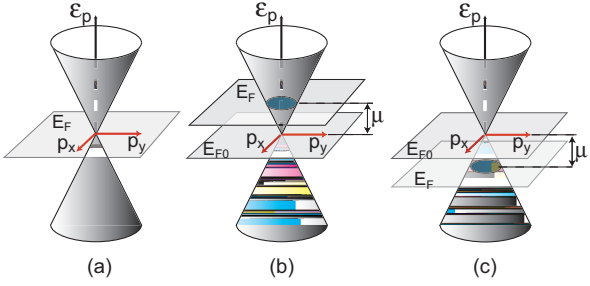}}
\caption{(Color online) Conical energy spectrum. (a) Undoped, (b)
electron-doped, and (c) hole-doped spectrum} \label{fig-spectrum}
\end{figure}

Graphite attracts attention of experimentalists and theorists for
a long time. The interest is explained by unusual properties of
this quasi-two dimensional material, which are mostly related with
the existence of the Dirac or conic point in the electronic
spectrum (see Fig. \ref{fig-spectrum}). Though the theory has
predicted the existence of such point in graphite many decades
ago\cite{Wal}, only recently the experimental  evidences of its
existence were received: first in graphite\cite{DirakP}, which is
believed to be a stack of weakly coupled atomic layers, and soon
after it in graphene\cite{Novoselov05,Kim}. The latter discovery
has triggered an avalanche of experimental and theoretical works.
Moreover, graphene can display unusual properties as a part of
normal-superconducting hybrid structures: For example, the Andreev
reflection has been predicted to have new features not
characteristic for typical contacts\cite{Beenakker06}.

Thorough investigation of graphite has revealed also evidences of
intrinsic superconductivity in doped samples (see Refs.
\cite{Kopelevich01,Kopelevich06} and references therein). Various
mechanisms of superconductivity in graphene have been considered
theoretically. Phonon and plasmon mediated mechanisms were
discussed in Ref. \cite{Uchoa07} whereas a resonating valence bond
model was proposed in Ref. \cite{Doniach07}.  The Cooper pairing
in the undoped graphene may experience problems because the Fermi
surface shrinks near the Dirac point and reduces to zero the
number of states at the Fermi energy. Indeed, it was shown within
the BCS model\cite{Marino} that the superconducting transition in
the undoped graphene possesses a quantum critical point at a
finite interaction strength below which the critical temperature
vanishes. However, one would expect that the electrons in graphene
may become unstable towards formation of Cooper pairs for any
finite pairing interaction if doping shifts the Fermi level away
from the Dirac point because the behavior of electrons in the
latter case bears more resemblance to that in usual metals. This
idea has been qualitatively discussed in Refs.
\cite{Uchoa07,Zhao06} and verified within the resonating valence
bond model in Ref. \cite{Doniach07}.

The aforementioned investigations of superconductivity in graphene
or graphite (except for Ref. \onlinecite{Marino}, where only the
undoped case was considered) were done taking into account the
specific details of each particular pairing mechanism. However, it
would be worthwhile to perform the analysis in a more general form
independent of a particular nature of the pairing mechanism. In
the present Letter we apply the standard $s$-wave BCS model for
the Dirac spectrum of electrons with a minimum number of
parameters characterizing the pairing interaction, i.e., its
intensity and the range of interaction in the momentum space. The
values of these two parameters may vary depending on the
mechanism. Such approach inevitably ignores some details and thus
is less accurate. However, we hope that the loss of accuracy is
compensated by a more general and transparent picture of the most
essential features of the Cooper pairing in the systems with the
Dirac spectrum.

In what follows we calculate the critical temperature, the
superconducting energy gap, and the supercurrent as functions of
the doping level and the pairing interaction strength. Without
doping the critical temperature vanishes below some finite value
of the interaction strength. However, the critical temperature is
nonzero for any nonzero electron or hole doping level when the
Fermi energy is shifted from the Dirac point of the normal-state
electronic spectrum. This provides the quantitative basis for the
earlier conjectures of Refs.~\cite{Uchoa07,Zhao06} and agrees
qualitatively with the results of Ref.~\cite{Doniach07} for
resonating valence bond model. Moreover, by analyzing the effect
of the Dirac point on the supercurrent we demonstrate a novel
feature that, as distinct from the usual superconductors, the
supercurrent density is not proportional to the total number of
electrons but is drastically decreased due to the presence of the
Dirac point. Finally, we estimate characteristic length scales
(penetration depth and coherence length), relevant for
determination of the critical magnetic fields.

Consideration of a two-dimensional model requires a few comments
concerning the applicability of the mean-field approach. It is
well known that the superfluid transition in a two-dimensional
system occurs in the form of the Berezinskii-Kosterlitz-Thouless
transition at a temperature lower than the mean-field transition
temperature. Therefore our calculations provide the upper bound on
the critical temperature and give a good estimate for the
temperature scale of the transition \cite{Doniach07}. Moreover,
the applicability of the mean-field approach improves for graphite
where a nonzero interplanar coupling, however small, does always
exist.

\paragraph{Spectrum.}--
We assume the energy spectrum in graphene in the form
\[
\epsilon_{\bf p} =\pm v \sqrt{p_x^2+ p_y^2}+E_{F0}\ .
\]
The upper or lower sign refers to the conduction or valence band,
respectively; $E_{F0}$ is the Fermi energy without doping when the
Dirac point lies at the Fermi level. If the Fermi energy  is
shifted by some amount $\mu$ due to doping, $E_F=E_{F0}+\mu $,
(see Fig. \ref{fig-spectrum}) the energy measured from the Fermi
level is
\[
\xi_{\bf p}=\epsilon_{\bf p} -E_F =\pm v p -\mu \ .
\]
The group velocity is $ d\xi_{\bf p}/d{\bf p}=\pm v {\bf n} $
where $ {\bf n}={\bf p}/p$. For electron doping, $\mu >0$, we have
for $\xi_{\bf p}<0$
\begin{equation}
p=\left\{ \begin{array}{lr} -(\xi_{\bf p} +\mu)/v\ , & \xi_{\bf p}<-\mu \\
(\xi_{\bf p} +\mu)/v\ , &
 -\mu <\xi_{\bf p}
\end{array}\right. \ . \label{dp-e}
\end{equation}
Similar
relations take place in the case of hole doping, $\mu =-|\mu|$, as
well.

\paragraph{BCS gap equation.}--
We use the standard BCS theory and assume an s-wave pairing
interaction $V_p=-|V_p|$,  where $|V_p|\sim |V| a^2$ is the
Fourier transform of the pairing potential, $V$ is the energy
amplitude, and $a$ is the range of interaction. We do not consider
here the nature of pairing interaction, but refer the reader to
Refs. \cite{Uchoa07,Zhao06,Sasaki07,Doniach07} where various
possible mechanisms are discussed. The coupling constant $\lambda$
is introduced through the equation
\[
\frac{|V_p|}{2\pi\hbar^2 v^2}  =\left\{ \begin{array}{lr}
\lambda /\xi_m \ , & |\xi_{\bf p}| <\xi_m \\
0 \ , & |\xi_{\bf p}|>\xi _m \end{array}\right. \ .
\]
Here $\xi_m$ determines the interval where attractive interaction
is present. In what follows we consider the case of low doping
when $|\mu| <\xi_m$.

The BCS gap equation in a spatially uniform case is
\begin{equation}
1= \frac{1}{2} \int |V_p| \, \frac{d^2 p}{(2\pi\hbar)^2}
\frac{1}{E_{\bf p}}[1-2n(E_{\bf p})] ,\label{BCSgapeq}
\end{equation}
where the energy of excitations is $ E_{\bf p}= \sqrt{\xi_{\bf
p}^2+\Delta ^2}$,  the phase volume is $ d^2 p =p \, d p\, d\phi
$, where $\phi$ is the azimuthal angle of ${\bf n}$, and $n(E_{\bf
p})$ is the equilibrium Fermi distribution of quasiparticles with
energies $E_{\bf p}$. For zero temperature $1-2n(E)={\rm
sign}(E)$. With help of Eq. (\ref{dp-e}), the BCS gap equation
becomes
\begin{equation}
\frac{\xi_m}{\lambda}=\sqrt{\xi_m^2+\Delta_0^2}-\sqrt{\mu^2+\Delta_0^2}
+|\mu| \ln \! \left[\frac{|\mu|\! +\!\sqrt{\mu
^2+\Delta_0^2}}{\Delta_0 }\right] \label{gap-gen}
\end{equation}
for both electron and hole doping.

For zero doping $\mu =0$ we have
\begin{equation}
\Delta_0 =\xi_m (\lambda^2-1)/2\lambda \ . \label{eqDelta-zero2}
\end{equation}
Nonzero $\Delta$ is possible only for the strong-coupling limit
$\lambda >1$ \cite{Marino}. However, Eq. (\ref{gap-gen}) shows
that, for a finite doping, a finite $\Delta_0$  exists even in the
weak coupling limit $ \lambda < 1 $. In the case of low doping
level when $\Delta_0$ is small, Eq. (\ref{gap-gen}) gives the gap
in a BCS form
\begin{equation}
\Delta_0 =2|\mu |
\left(-\frac{\xi_m}{|\mu|}\frac{1-\lambda}{\lambda} -1\right)\ .
\label{Delta0}
\end{equation}
with the prefactor determined by the doping level $|\mu|$ rather
than by the range of interaction.

\begin{figure}[t]
\centerline{\includegraphics[width=0.8\linewidth]{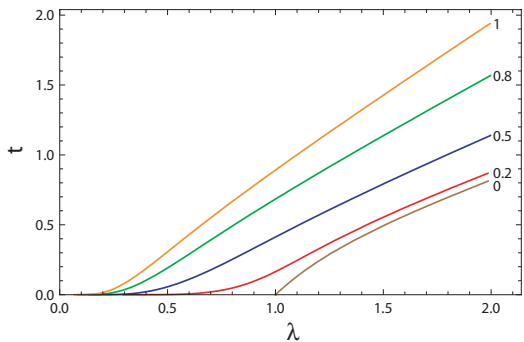}}
\caption{(Color online) Normalized critical temperature
$t=2T_c/\xi_m$ as a function of the interaction constant $\lambda
$ for various doping levels $|\mu|/ \xi_m$. The quantum critical
point is at $\lambda =1$ and $|\mu|=0$.} \label{fig-Tc}
\end{figure}

\paragraph{Temperature dependence.}--
For a finite temperature Eq. (\ref{BCSgapeq}) yields the gap
equation
\begin{eqnarray}
\frac{\xi_m}{\lambda}&=&2T\ln\left[\frac{\cosh(\sqrt{\xi_m^2+\Delta
^2}/2T)}{\cosh (\sqrt{\mu^2+\Delta ^2}/2T)}\right]\nonumber \\
&&+|\mu| \int_{0}^{|\mu|} \tanh\frac{\sqrt{\xi^2+\Delta ^2}}{2T}\,
\frac{ d\xi}{\sqrt{\xi^2+\Delta ^2}}\ . \label{BCS-i-T}
\end{eqnarray}
For $T\rightarrow 0$ we return to Eq. (\ref{gap-gen}). Equation
(\ref{BCS-i-T}) leads to the equation for the critical temperature
\begin{equation}
\Phi\left(\xi_m/2T_c ; \lambda\right)=F\left( |\mu|/2T_c\right)
\label{BCS-i-Tc}
\end{equation}
where
\begin{eqnarray*}
\Phi\left(y; \lambda\right)&=&\lambda^{-1}y -\ln (\cosh y)\\
F(x)&=&x\int_0^x (x^\prime)^{-1}\tanh x^\prime \, dx^\prime - \ln
(\cosh x)
\end{eqnarray*}
where $F(x)>0$. The critical temperature found from Eq.
(\ref{BCS-i-Tc}) is plotted in Fig. \ref{fig-Tc}.

For $\mu =0$ the critical temperature satisfies
$\Phi\left(\xi_m/2T_c ; \lambda\right)=0$, i.e.,
\begin{equation}
\xi_m/\lambda =2T_c\ln\left[\cosh(\xi_m/2T_c)\right]\ .
\label{BCS-Tc}
\end{equation}
This equation has a solution only for interaction strength above
the quantum critical point, $\lambda
>1$ (see Fig. \ref{fig-Tc}). If $\lambda \rightarrow 1$ we have
$ T_c= \xi_m(\lambda -1)/2 \ln 2$ which vanishes at $\lambda =1$.
Comparing this with Eq. (\ref{eqDelta-zero2}) we find $
\Delta_0=T_c2\ln 2 $. In the other limit $\lambda \gg 1$ we find
$T_c=\xi_m \lambda /4$ and $\Delta_0=2T_c$. These results agree
with Ref.\cite{Marino} where only undoped case was considered.

However, for any low but finite doping level the critical
temperature is finite. Consider weak coupling limit $\lambda \ll
1$ where we expect $T_c \ll \mu$. Indeed, the l.h.s. of Eq.
(\ref{BCS-i-Tc}) is $\Phi(y;\lambda)=y(\lambda^{-1}-1)+\ln 2$
already for $T_c \ll \xi_m$. On the other hand, for $x\gg 1$ the
r.h.s. of Eq. (\ref{BCS-i-Tc}) is
\begin{equation}
F(|\mu|/2T_c)=\frac{|\mu|}{2T_c }\ln  \left[\frac{2|\mu|
\gamma}{e\pi T_c}\right]+\ln 2  \ . \label{rhs-lowmu}
\end{equation}
where $ \gamma =e^C=1.78 $ and $C=0.5772$ is the Euler constant.
This yields
\begin{equation}
T_c=\frac{2|\mu| \gamma}{\pi }\exp\left[ -\frac{\xi_m
(1-\lambda)}{\mu \lambda}-1\right] \ , \label{Tc}
\end{equation}
resulting in the BCS relation $ \Delta_0=(\pi /\gamma)T_c =1.76
T_c $.

Consider the vicinity of the quantum critical point $\mu =0$ and
$\lambda =1$. On the weak coupling side $\lambda < 1$, the
critical temperature is given by Eq. (\ref{Tc}) which is exact
provided $T_c\ll |\mu|$, i.e., for $1-\lambda\gg |\mu|/\xi_m$. For
$|\mu|/\xi_m\sim 1$ and $\lambda \rightarrow 1$, Eq. (\ref{Tc})
works also reasonably well. For example, Eq. (\ref{Tc}) gives $T_c
\approx 0.42|\mu|$ for $\lambda =1$. This can be compared to the
exact value for $\lambda =1$ which is found from the condition
$F(|\mu|/2T_c)=\ln 2$ resulting in $T_c\approx 0.40 |\mu|$. In the
limit $\mu \ll T_c\ll \xi_m$ which is more appropriate on the
strong-coupling side of the quantum critical point
\[
T_c=\frac{\xi_m(\lambda-1)+\sqrt{\xi_m^2 (\lambda -1)^2+\mu^2\,
2\ln 2}} {4\ln 2}\ .
\]
This holds for $|\mu|/\xi_m\ll \lambda -1 \ll 1 $, but also
matches with the exact $T_c$ by the order of magnitude when
$\lambda \rightarrow 1$.

Therefore we come to the conclusion that a finite $T_c$ does
always exist for a finite $\mu$. If $\lambda \gtrsim 1$, the
critical temperature is close to that determined by Eq.
(\ref{BCS-Tc}) as long as $\mu \ll \xi_m$. If $\lambda \lesssim 1$
we essentially have Eq. (\ref{Tc}).

\paragraph{Supercurrent.}--
Let us assume a homogeneous flow of the condensate: $\Delta
=|\Delta|e^{i{\bf k}_s{\bf r}}$, where ${\bf k}_s=\nabla \chi $ is
a constant gradient of the order-parameter phase. In the presence
of magnetic field, ${\bf k}_s=\nabla \chi-(2e/\hbar c){\bf A}$.
Consider the state described by the particle-like and hole-like
Bogoliubov-de Gennes wave functions
\[
u({\bf r})=u_{\bf p}e^{i{\bf p}_+\cdot{\bf r}/\hbar}\ , \; v({\bf
r})=v_{\bf p}e^{i{\bf p}_-\cdot {\bf r}/\hbar},
\]
where ${\bf p}_\pm ={\bf p}\pm \hbar {\bf k}_s/2$,
\[
E_{\bf p}=E_D+E_{\bf p}^{(0)}\ , \; E_{\bf p}^{(0)}=
\sqrt{\xi_{\bf p}^2+|\Delta|^2}\ ,
\]
$E_D=(d\xi_p /d{\bf p})\hbar {\bf k}_s/2$ is the Doppler energy,
and
\[
u_{\bf p}=\frac{1}{\sqrt{2}} (1+\xi_{\bf p}/E^{(0)}_{\bf
p})^{1/2}\ , \; v_{\bf p}=\frac{1}{\sqrt{2}} (1-\xi_{\bf
p}/E^{(0)}_{\bf p})^{1/2}
\]
are the coherence factors. The standard expression for the current
is
\begin{equation}
{\bf j}= 2e\sum _{\bf p} \left[\frac{\partial \xi_{{\bf
p}_+}}{\partial {\bf p}}|u_{\bf p}|^2 n(E_{\bf p}) -\frac{\partial
\xi_{{\bf p}_-}}{\partial {\bf p}}|v_{\bf p}|^2[1-n(E_{\bf
p})]\right] \ . \label{BdGcurrent}
\end{equation}

Expanding Eq. (\ref{BdGcurrent}) for small $E_D\ll \Delta, T$ and
making shift in the momentum variable we find for the
two-dimensional current density in the linear response regime
\begin{eqnarray*}
{\bf j} =e\! \int \!\! \frac{d^2 p}{4\pi^2\hbar } \frac{\partial
\xi_{\bf p}}{\partial {\bf p}}\! \left(\frac{\partial \xi_{\bf
p}}{\partial {\bf p}}\cdot {\bf
k}_s\right)\frac{\partial}{\partial \xi_{\bf p}}\left[
\frac{\xi_{\bf p}}{2E^{(0)}_{\bf p}}[1-2n(E^{(0)} _{\bf p})]\right]\\
+2e \int \frac{d^2 p}{4\pi^2\hbar^2} \frac{\partial \xi_{\bf
p}}{\partial {\bf p}}\left[ n(E_{\bf p})- n(E^{(0)}_{\bf
p})\right]\ .
\end{eqnarray*}
This yields the  current
\[
{\bf j} =(e\Lambda /4\pi \hbar){\bf k}_s
\]
where $\Lambda$ is the characteristic energy. For zero temperature
we have
\[
\Lambda=2|\Delta|+\frac{\mu^2}{\sqrt{\mu^2+|\Delta|^2}}-
\frac{|\Delta|^2 }{\sqrt{\mu^2+|\Delta|^2}}\ .
\]
In contrast to the usual superconductors the supercurrent density
is not proportional to the total electron density,  being
drastically affected by the presence of the Dirac point. In
particular, for weak coupling limit, $|\Delta|\ll \mu$, the
current $ {\bf j} =e\mu {\bf k}_s/ 4\pi\hbar $ is proportional to
$|\mu| \propto \sqrt{n}$, where $n$ is the density of free
carriers provided by doping. Near the quantum critical point when
$T\ll \xi_m$, the current is determined by the superconducting gap
itself. Indeed, for zero doping, Eq. (\ref{BdGcurrent}) yields
\[
\Lambda=|\Delta|\tanh \frac{|\Delta|}{2T}\ .
\]
For low temperatures, $T\ll |\Delta|$, we have $ {\bf
j}=e|\Delta|{\bf k}_s/4\pi\hbar $. Close to $T_c$, where $|\Delta|\ll T_c$, the
current assumes the Ginzburg-Landau form $ {\bf j}= e|\Delta|^2
{\bf k}_s/ 8\pi\hbar T_c$.

\paragraph{Characteristic lengths and critical fields.}--
As usual the critical fields are determined by two spatial scales: the coherence
length $\xi_0$ and the London penetration depth $\lambda_L$.
At zero temperature the London penetration length for a graphene layer
with thickness $d$ is
\[
\lambda_L^{-2}=\frac{2e^2\Lambda }{\hbar^2 c^2 d}\ .
\]
It diverges near the quantum critical point $\lambda \rightarrow
1$, $\mu \rightarrow 0$. For the undoped case the London length is
$\lambda_L =(\Phi_0/\pi) \sqrt{d/2\Delta}$, where $\Phi_0=\pi
\hbar c/e $ is the magnetic-flux quantum. Close to the critical
temperature the London length, $\lambda_L =(\Phi_0/\pi\Delta)
\sqrt{T_cd}$, is inversely proportional $\sqrt{T_c-T}$ as in
conventional superconductors. The coherence length has a standard
form: $\xi_0\sim \hbar v_F/\Delta$. Thus the Ginzburg-Landau
parameter $\kappa =\lambda_L/\xi_0$, which characterizes the type
of superconductivity, does not depend on the temperature near
$T_c$ as is the case in conventional superconductors:
\[
{\lambda_L\over \xi_0} \sim {c\over v_F}\sqrt{T_c d\over e^2}\ .
\]
For typical values $v_F=10^8$ cm/s,  $d=10^{-7}$ cm, and for
$T_c\sim 1$ K, the Ginzburg-Landau parameter is on the border
between the two types, $\kappa \sim 1$. Therefore, close to the
quantum critical point where $T_c \to 0$ the superconductivity
definitely becomes of type I.

To summarize, we have calculated the critical temperature, the
superconducting gap, and the supercurrent  as functions of the
doping level and of the interaction strength for an s-wave pairing
within the BCS model. The superconducting transition in the
undoped graphene has a quantum critical point with respect to the
interaction strength, which disappears for any finite doping level
such that a finite critical temperature exists for any weak
pairing interaction. The amplitude of the supercurrent  is
drastically affected by the presence of the Dirac point, which
leads to non-trivial behavior of the characteristic length scales
(penetration depth and coherence length) determining critical
magnetic fields.

\acknowledgements

We thank Y. Kopelevich and V. Eltsov for stimulating discussions.
This work was supported by the Forscheimer Foundation of the
Hebrew University of Jerusalem, by the Academy of Finland (grant
213496, Finnish Programme for Centers of Excellence in Research
2002-2007/2006-2011), by the ULTI program under EU contract
RITA-CT-2003-505313, and by the Russian Foundation for Basic
Research grant 06-02-16002.


\begin{thebibliography}{99}

\bibitem{Wal} P. R. Wallace, Phys. Rev. {\bf 71}, 622  (1947).
\bibitem{DirakP} I. A. Luk'yanchuk and Y.  Kopelevich,
Phys. Rev. Lett. {\bf 93}, 166402 (2004).

\bibitem{Novoselov05} K. S. Novoselov, A. K. Geim, S. V. Morozov, D. Jiang,
M. I. Katsnelson, I. V. Grigorieva, S. V. Dubonos, and A. A. Firsov,
Nature {\bf 438}, 197 (2005).

\bibitem{Kim}  Y. Zhang, Y.-W. Tan, H. L. Stormer, and P. Kim,
Nature {\bf 438}, 201 (2005).

\bibitem{Beenakker06} C.W.J. Beenakker, Phys. Rev. Lett. {\bf 97},
067007 (2006).

\bibitem{Kopelevich01} R. R. da Silva, J. H. S. Torres, and Y.
Kopelevich, Phys. Rev. Lett. {\bf 87}, 147001 (2001).

\bibitem{Kopelevich06} Y. Kopelevich, S. Moehlecke, and R. R. da Silva,
in {\it Carbon Based Magnetism}, edited by T. Makarova and F.
Palacio (Elsevier Science, 2006), Chap. 18.

\bibitem{Uchoa07} B. Uchoa and A. H. Castro Neto, Phys. Rev.
Lett. {\bf 98}, 146801 (2007).

\bibitem{Doniach07} A. M. Black-Schaffer and S. Doniach, Phys. Rev. B, {\bf
75}, 134512 (2007).

\bibitem{Marino} E. C. Marino, and Lizardo H.C.M. Nunes, Nuclear
Physics B, {\bf 741}, 404 (2006); Physica C {\bf 460-462}, 1101
(2007); Nuclear Physics B, {\bf 769}, 275 (2007).


\bibitem{Zhao06} E. Zhao and A. Paramekanti, Phys. Rev.
Lett. {\bf 97}, 230404 (2006).


\bibitem{Sasaki07} K. Sasaki, J. Jiang, R. Saito, S. Onari, and Y.
Tanaka,   J. Phys. Soc. Jpn. {\bf 76}, 033702 (2007); arXiv:cond-mat/0611452.




\end{thebibliography}
\end{document}